\documentstyle[preprint,pre,aps]{revtex}

\tighten       

\begin{document}

\draft
\preprint{Version of \today }

\title{Chaotic motion of space charge wavefronts in
semiconductors under time-independent voltage bias}

\author{I.~R.\ Cantalapiedra$^1$ \cite{inma:email},
M.~J.\ Bergmann$^2$ \cite{mjb:email}, L.~L.\ Bonilla$^3$
\cite{bonilla:email}, and S.~W.\ Teitsworth$^2$
\cite{teitso:email} }

\address{$^1$Departamento de F\'{\i}sica Aplicada,
Universitat Polit\'{e}cnica de Catalunya, Dr.
Mara\~{n}\'{o}n 44, 08028 Barcelona, Spain\\
$^2$Department of Physics and Center for Nonlinear and
Complex Systems,  Box 90305, Duke University, Durham,
NC 27708-0305 \\
$^3$Departamento de Matem\'aticas, Escuela
Polit\'{e}cnica Superior, Universidad Carlos III de
Madrid, \\ Avenida de la Universidad 30; 28911
Legan\'{e}s, Spain}

\date{\today}

\maketitle

\begin{abstract}
A standard drift-diffusion model of space charge wave
propagation in  semiconductors has been studied
numerically and analytically under dc  voltage bias.
For sufficiently long samples, appropriate contact
resistivity and applied voltage
-  such that the sample is biased in a regime of
negative differential resistance - we find chaos in the propagation of
nonlinear fronts (charge monopoles of alternating sign)
of electric field.  The chaos  is always
low-dimensional, but has a complex spatial structure;
this behavior can be interpreted using a finite
dimensional asymptotic model in which the front
(charge monopole) positions and the electrical current
are the only dynamical variables.
\end{abstract}

\narrowtext

\section{Introduction}
\label{intro}

The dynamics of propagating solitary waves (pulses) and
monotone fronts in nonlinear partial differential
equations ({\em pde}'s) have been the focus of
extensive  research.  Such problems are of interest in
a wide range of fields including biology (population
dynamics) \cite{mur90}, chemical reactions and
combustion \cite{kur84,wil85}, plasma physics
\cite{hort99}, and semiconductor electronic transport
\cite{nie95}.  A common problem is to  understand and
predict the form and speed of the waves, as well as
the  possibility of multiple excitations in a sample of
finite or infinite extent, and their interactions.  We
focus here on a model of electrical  conduction in
extrinsic semiconductors (involving time and only one
spatial  dimension) which exhibits negative
differential resistance (NDR) and moving domains of
high electric field. The model is specifically relevant
to experiments on cooled bulk p-type Ge under voltage
bias conditions, \cite{teitsPRL83,kahnPRB91}, but much of the
observed qualitative behavior
applies to a broad class of semiconductor systems with space
charge instabilities.  Phenomena observed for the p-Ge
system include time-periodic oscillation of the
current in a purely resistive external circuit under
dc  voltage bias due to the periodic creation of a
solitary wave at the  injecting contact, its motion
inside the semiconductor and its annihilation  at the
receiving contact \cite{kahnPRB91}. There is some
similarity to the  Gunn effect in n-GaAs, except that:
(i) the local current density versus field
characteristics [see $j(E)$ below] in p-Ge presents an
increasing third branch after the NDR region, and (ii)
the solitary waves in p-Ge move much more slowly than
the carrier drift velocity (the case in the usual Gunn
effect) due to the generation-recombination dynamics of
ionized traps  which dominate the transport properties.
Other experimental observations  include intermittency
near the onset of the oscillatory instability
\cite{kahnPRB92,kahn45PRB92}, and ``spatiotemporal''
chaos under combined dc and ac voltage bias \cite{kahnPRL92}.
Another important feature is the integral constraint which
corresponds to voltage bias applied across sample contacts 
in semiconductor problems.  Integral constraints also occur in
other situations, for example,
expressing mass conservation in problems of phase separation in
binary mixtures and in certain biological problems
\cite{mur90,rub92,rey94}.

Many of these phenomena have been successfully
explained by means of a drift-diffusion model which
includes impurity trapping of mobile holes and  impact
ionization of neutral acceptors
\cite{teitsPRL83,wesJAP85}. Although  much work has
been done on this model problem (see \cite{bergPRB96}
and references therein), important basic questions
concerning its asymptotic description and chaos under
time independent voltage bias are still open. In this
paper, we present numerical simulation results which show
chaos under dc voltage bias associated with multiple
shedding of wavefronts. Multiple shedding of wavefronts
occurs for appropriate values of contact resistivity, and was
recently predicted on the basis of asymptotic calculations 
\cite{bonPD97}. Here we introduce a finite dimensional
model which provides a simplified description of space
charge wave dynamics in long samples. This model uses
relevant information from the asymptotics in
Ref.~\cite{bonPD97}, although we do not rigorously
derive it from such asymptotic calculations.
Nevertheless, solutions of the simplified model are in good
agreement with the results of direct numerical
simulations.

The paper is organized as follows. In Section II we
present a drift-diffusion model which
accurately describes the propagation of space charge
wavefronts in extrinsic semiconductors under dc voltage
bias such as p-Ge. In Section III we present 
numerical simulation results of the drift-diffusion model
which indicate that wavefronts may propagate chaotically 
due to multiple shedding of wavefronts for appropriate bias in sufficiently long
samples. Section IV presents the asymptotic model and
numerical simulations thereof which are then compared
with results from the drift-diffusion model.
Conclusions are finally presented in Section V.

\section{The reduced drift-diffusion model}
\label{drift-diff}

In dimensionless form, the drift-diffusion model
equations for a sample of length $L$ can be written in
a form \cite{bonPD91,bonPRB92}:

\begin{eqnarray}
\frac{\partial^{2}{E}}{\partial{x} \partial{t}} + J\,
\frac{K+R}{V^{2}}\,\left(\frac{V'}{K+R}\,
\frac{\partial E}{\partial t} \right.\nonumber\\
\left. + V\,
\frac{\partial E}{\partial x} + j(E) - J\right) =
\frac{1}{V}\,\frac{dJ}{dt},\label{red}\\ {1\over
L}\,\int_0^L E(x,t)\ dx = \phi,
\label{gc}\\ E(0,t) = \rho J(t). \label{bc}
\end{eqnarray}

The first equation describes the spatio-temporal
evolution of the  electric field $E(x,t)$ inside the
sample, where $J(t)$ is the total current density.  The
transport coefficients $V$, $K$ and $R$ are,
respectively, the average (drift) velocity of
microscopic charge carriers (holes in the case of
p-Ge), and coefficients describing the creation of free carriers via
impurity impact ionization and the 
destruction of free carriers by capture onto an
available impurity trapping site (i.e., neutral
acceptor).  All are nonlinear functions of electric
field, and their forms and plots have been
discussed  extensively in the literature
\cite{nie95,wesJAP85,bergPRB96}.  In this paper we use
the same forms as in Ref. \cite{bergPRB96}.  Equation (\ref{gc})
is a global constraint which expresses the voltage bias
condition, and Eq. (\ref{bc}) is a boundary condition
which represents the Ohmic injecting
contact at $x=0$ with contact resistivity $\rho>0$.
We refer to  Eqs. (\ref{red}) - (\ref{bc}) as the {\it
reduced} drift-diffusion model because they are derived
from a full drift-diffusion model by systematically
dropping terms that correspond to short length and time
scale processes of diffusion and displacement current,
respectively.  For a precise derivation as well as a
complete table of conversion factors to dimensional
units, see \cite{nie95} and \cite{bonPD91}.
Some of them are: time 2.1 $10^{-3}$ ms, length 0.01 mm,
electric field 10 V/cm, density current 128.16 $\rm{mA/cm^2}$,
cross-sectional area 0.16 $\rm{cm^2}$. 
We have adopted the same symbol for both dimensional and
non-dimensional variables.

The qualitative nature of much of the dynamical
behavior found in the reduced drift-diffusion model---e.g., 
the instability of the stationary 
electric-field profile and propagating
high-field domains
\cite{bonPRB92,canPRB93,bonSST94}---depends only on the
presence of a region of  negative slope of the
homogeneous stationary current density $j(E) = V(E)\,
\{\alpha K(E)/[K(E) +R(E)] - 1 \}$,
\cite{bergPRB96,bonPRB92,canPRB93} over an interval of
positive fields, and not on the exact form of the
underlying coefficients. This is particularly true when
the sample is closely  compensated (the ratio of the
acceptor concentration to the donor  concentration,
$\alpha$, is only slightly larger than 1). Then $j(E)$
is N-shaped for large enough positive fields: there is
an interval $(E_M,E_m)$ between the abcissas of the
maximum [$j(E_M) = j_M$] and the minimum [$j(E_m) =
j_m > 0$] current density for which $dj/dE<0$ and $j(E)>0$,
as shown in Fig.~\ref{j_sh}.  Also shown in
Fig.~\ref{j_sh} is the injecting contact
characteristic which plays a crucial role in
determining when new fronts are injected into the
sample. The critical current density, $J_c$,
corresponds to the intersection of the contact
characteristic with the homogeneous stationary current
density. The role of $J_c$ has been elucidated using 
a rigorous asymptotic analysis of the system
Eqs. (\ref{red}) - (\ref{bc}) in the limit as $L\to \infty$
\cite{bonPD97}, and is discussed further in Section IV.

\section{Numerical results}

To solve the system of equations (\ref{red}) -
(\ref{bc}) for $E$ and $J$, we discretize the
equations using finite difference approximations to the
derivatives and employ an implicit method to generate
the solution. The initial condition for the electric
field is spatially uniform with a value that is
consistent with the global constraint, Eq.~(\ref{gc}).
In Fig.~\ref{Ext.p1} we show a space-time plot of the
electric field and associated current density for $\phi
= 6.25$ V/cm, just above the threshold voltage value for which
propagating domain behavior occurs. The gray scale ranges
from 5.3 V/cm (black) to 14.2 V/cm (white), and this scale is 
used in all similar plots that follow. The
dimensionless sample length is 3800 corresponding to a real p-Ge
sample of length 3.87 cm, and the contact resistivity
$\rho$ is 780 $\rm{\Omega cm}$ corresponding to a value
of 10.0 in dimensionless units. This case corresponds well
to experimental data, but published data in p-Ge were
only presented for one sample and relatively low bias
values and contact resistivity.  We clearly see  that a
single domain moves across the sample  at constant
speed, until it reaches the receiving contact. As it
disappears, a new wave is created at the injecting
contact and the process repeats periodically. The
current versus time plot indicates that the current is
steady when the domain moves in the sample interior,
while there is an increase when the domain reaches the
receiving  contact.  It is  important to note that the
fronts of changing electric field (or equivalently, the
regions of nonzero charge density) are sharp in space
relative to other physical length scales for this
problem, i.e, the extent of the flat top domains  or
the sample length. We have found that this separation
of length scales increases with sample length and holds
for most biases of interest.  It is only for voltages
near the onset point that  one tends to observe rounded
solitary waves  rather than well-separated pairs of
fronts; this is the dominant space charge wave
structure observed in shorter samples and has been
extensively reported
\cite{bergPRB96,bonPRB92,canPRB93}.

As the bias increases, the propagating domain becomes
fatter and eventually  a second small domain is
nucleated and propagates part way into the sample; but
it dies before reaching the receiving contact or
merging with the larger domain.  At even larger bias
values the second domain merges with the primary domain
near the receiving contact and this situation is shown
in Fig.~\ref{Ext.p2} which corresponds to $\phi = 7.25$
V/cm. Again the current is plotted on the far right of
the figure.  When the first domain reaches the
receiving contact the current increases.  Instead of
immediately starting the nucleation of a new wave, the
area lost by the dying wave is gained by the trailing
wave -- note that the width of the trailing wave
increases after the leading wave starts to disappear.
The current increases, reaching a local maximum just
before the trailing domain touches the leading domain,
that is, the fronts collide. The current increases
abruptly after the front collision and it rises to a
global maximum at which point a new domain begins to
nucleate at the injecting contact. As the domain forms,
the current decreases and reaches a minimum at which
point a new domain detaches and begins to propagate.
Then the current increases until a second smaller
domain is nucleated. Finally, the current settles to a
rather low constant level as the two domains move
steadily and in unison across the interior region of
the sample. Current behavior is apparently dominated by
the major events involving the fronts: collisions  with
the contacts or with each other. This suggests the
viability of a dynamical model that focuses on discrete
front motions and the current $J(t)$.

At larger biases, the portion of the sample occupied by
the high field value $E_3$ is larger, reducing the
separation between domains, and giving more complicated
$E(x,t)$ structure and $J-t$ behavior.   In
Fig.~\ref{Ext.ch1} we show a space-time plot and
current for what appears to be a chaotic state for an
applied bias of $\phi = 10.0$ V/cm.  The spatiotemporal
dynamics possess a great deal of  structure and
complexity.  The process of multiple domain shedding is
similar to that for the previous case.  The large
current peaks correspond to nucleation of leading
domains. The leading domains cross the sample without
catching up or forward-colliding with any other  high
field regions, and are indicated by dark regions that
extend all the way across the space-time diagram.  Note
also the larger spatial extent of the leading domains
than in Figs. ~\ref{Ext.p1} and ~\ref{Ext.p2}.  The
aperiodicity of the current is  reflected in the
irregular appearance of the maximum current peaks or,
equivalently, of the dark strips that extend across the
entire sample. In between them are a number of local
maxima corresponding to the shedding of trailing
domains.

Figure ~\ref{phase} (a) shows a bifurcation diagram in
which we plot all values of successive current maxima
as a function of $\phi$.  An important feature in this
diagram is the apparent presence of windows of chaotic
behavior with a large number of points being visited.
For periodic states we see a small number of points
corresponding to perfectly repeating current maxima.
Also, in the periodic regimes there are points where
various branches merge or disappear and these
correspond to the development or destruction of
trailing domains.  For the parameter values selected
here we do not observe period doubling.  We conjecture
that the route to chaos here is of boundary crisis type
in which the attractor collides with a periodic orbit
on its basin boundary \cite{Ott}.

In Fig.~\ref{phase} (b) we show the largest Lyapunov
exponent $\lambda_1$ versus $\phi$ for the reduced
drift-diffusion model. This unambiguously confirms the
presence of chaos in the ``chaotic'' windows.  The next
two exponents have been calculated and are never
positive, so that the chaos we see is of a
low-dimensional variety. To compute the exponents, we
used an algorithm outlined in Ref. \cite {arker-chua}
adapted for use with partial differential equations and
using adaptive control of the integration time step
\cite {berg96thesis}. The values of $\lambda_1$ are
zero in the periodic regimes as they should be for
periodic behavior.  The smallness of $\lambda_1$ in the
chaotic regime is easily understood by recalling the
period of the system, about 1000 non-dimensional time
units. This indicates that the chaos originates in
processes that occur on time scales on the order of
the  front transit time across the sample.  To our
knowledge, this is the first time that chaos due to
multiple shedding of wavefronts has been observed in a
drift-diffusion model of this type.  This type of
chaotic behavior  has not been reported for experiments
on p-Ge with time-independent voltage bias, most
likely because experimentally studied samples were too
short, biases were not sufficiently large, or contact
resistivity was too low. However, in early
experimental studies of the Gunn effect in GaAs, J. B.
Gunn \cite{gun64}  observed that for long samples
current oscillations were almost completely random,
resembling white noise.  He also found that short samples produce
aperiodic oscillations when circuit impedance is
sufficiently large.  It is plausible that Gunn may have
observed a similar form of chaos to the one we have
found numerically. In the next section we develop an
asymptotic model in which the chaotic dynamics  is
understood to arise from the aperiodic nucleation of
fronts at the injecting contact.

\section{Asymptotic model}
The asymptotic model used in this paper consists of
describing the evolution of the current when
all wavefronts are detached
from the injecting contact by an appropriate
ordinary differential equation ({\em ode}) for $J$,
tracking the position of the wavefronts and proposing a
simplified mechanism for creation and destruction of
wavefronts. While our new model is compatible with the
asymptotic calculations of Ref.~\cite{bonPD97}, we have
not rigorously derived it from these calculations.
Instead, we have proposed a simplified dynamics to
account for our numerical observations motivated by
asymptotic results.

We begin by assuming that $E_M<\phi<E_m$ in
Fig.~\ref{j_sh} and that $E_M/j_M <\rho < E_m/j_m$.
Then a Gunn effect mediated by solitary waves occurs
\cite{bonPD92}, as shown by the numerical simulation in
Fig.~\ref{Ext.p1}. For appropriate parameter values,
there appears a regular oscillation of the current
caused by repeated creation, motion and destruction of
high-field domains in the sample. High-field domains
are formed by two wavefronts separating a region where
the electric field is uniform and large from regions of
uniform low field. Clearly, there are positively and
negatively charged wavefronts, having $\partial
E/\partial x >0$ or $\partial E/\partial x <0$,
respectively. Near the contacts, there are narrow
boundary layer regions where the electric field changes
abruptly. Creation of high-field domains occurs at the
injecting contact, via an instability of the boundary
layer which expels a high-field domain from the
injecting contact to the bulk of the sample. Typically,
the total current changes most during wavefront
creation and destruction events. In the limit as
$L\to\infty$, space and time scales are $x/L$ and
$t/L$, respectively \cite{bonPD97}. Then $j(E)=J$,
except in wavefronts and boundary layers at the
contacts. If the field profile consists of a single
high-field domain detached from the contacts, we have
$E=E_3(J)$ inside the domain and $E = E_1(J)$, outside,
where $E_1<E_2<E_3$ are the three zeros of $j(E)-J$ for
$j_m<J< j_M$, \cite{bonPD97}. High and low-field
regions are joined by wavefronts which are the unique
solution of Eq.\ (\ref{red}) (with zero right hand side) in
the moving coordinate $\chi = x- X_{\pm}(t)$,
$dX_{\pm}/dt = c_{\pm}(J)$ (the signs + or -- refer to
the charge inside the wavefront) and appropriate
boundary conditions. For example, at a positively
charged wavefront, $E\to E_1(J)$ as $\chi\to -\infty$
and $E\to E_3(J)$ as $\chi\to +\infty$. The numerically
determined values of $c_{\pm}(J)$ are shown in
Fig.~\ref{c(J)}.

\noindent Boundary layers obey (most of times) a
quasistationary version of Eq.\ (\ref{red}) with appropriate
boundary conditions on a semiinfinite spatial support.
The instantaneous value of the current $J(t/L)$
determines the field profile in the low and high
uniform-field regions and the velocity of the
wavefronts.

Next, assume that we have an initial field profile
consisting of $N$ high-field domains (solitary waves),
each formed by two wavefronts located at $X_+^{(i)}(t)
< X_-^{(i)}(t)$. We shall number the wavefronts so that
$X_{\pm}^{(i)}(t) > X_{\pm}^{(i+1)}(t)$, and if
necessary we shall consider $X_{-}^{(1)} = L$ and
$X_+^{(N)} = 0$. Then the positions $X_{\pm}^{(i)}(t)$
are given by
\begin{equation}
X_{\pm}^{(i)}(t) = \int_{t_{b,\pm}^{(i)}}^{t}
c_{\pm}(J(s))\, ds,   \label{X_pm}
\end{equation}
where $t_{b,\pm}^{(i)}$ denotes the time
at which the $i$-th monopole (with positive or negative
charge) was born at $x=0$.

The evolution of the total current density is
determined by the bias condition (\ref{gc}), which may
be approximately evaluated as
\begin{eqnarray}
 \phi = E_1(J) + [E_3(J)-E_1(J)]\, \sum_{i=1}^N
{X_-^{(i)} - X_+^{(i)}\over L} \label{asy1}
\end{eqnarray}
(terms of order $1/L$ and smaller have been ignored
here; note that the $X_{\pm}^{(i)}/L$ are of order
unity). We can get an {\em ode} for $J$ by
differentiating Eq.\ (\ref{asy1}) with respect to time and
then substituting $dX_{\pm}^{(i)}/dt = c_{\pm}(J)$ in
the result. We obtain
\begin{eqnarray}
{dJ\over dt} =  {1\over L}
\frac{(E_{3}-E_{1})^{2}}{{\phi-E_{1}\over j'_{3}}
+{E_{3}-\phi \over j'_{1}}}\, (n_+ c_+ - n_- c_-)\,
\label{asy2}\\
{dX_+^{(i)}\over dt} =  c_+(J),\quad
{dX_-^{(i)}\over dt} =  c_-(J),     \label{asy3}
\end{eqnarray}
where $i$ goes from 1 to $N$. The quantities $n_+$ and
$n_-$ are, respectively, the number of positive and
negative monopoles {\em detached} from the contacts
(i.e.,\ excluding possible monopoles at $x=0$ and
$x=L$), while $j'_{1}$ and $j'_{3}$ denote the derivative of the
static $j(E)$ characteristic with respect to electric
field, evaluated at $E_{1}$ and $E_{3}$, respectively.
 Notice that the system of equations
(\ref{X_pm}) to (\ref{asy3}) completely specifies the
behavior of current and field profile on the scales
$x/L$ and $t/L$, except that we do not have conditions
for determining when new fronts are emitted from the
injecting contact.

We start with the simple case of Fig.\ \ref{Ext.p1}:
the motion of a single high-field domain far from the
contacts. $J$ satisfies
Eq.\ (\ref{asy2}) with $n_+ = n_- = 1$, i.e., $dJ/ds =
A(J)\, [c_+(J) -c_-(J)]$, where $s=t/L$ and $A(J)>0$. As
shown in Figure \ref{c(J)}, $c_+(J)$ [resp.\ $c_-(J)$]
is a decreasing (resp.\ increasing) function of the
current. Therefore, $J$  evolves exponentially fast
toward the zero of the right hand side of this
equation, $J = J^{*}$. When the leading wavefront at
$x= X_-$ arrives at $x=L$, it disappears almost
instantaneously in the scale $s$, and we obtain $dJ/ds
= A(J)\, c_+(J) > 0$, so that the current increases.
The injecting boundary layer near $x=0$ ceases to be
quasistationary when $J$ surpasses the value $J_c$ at
which the line $J=E/\rho$ intersects the second
(decreasing) branch of $J=j(E)$. The precise
description of the instability, which results in
expelling a narrow high-field domain from $x=0$ to the
interior of the sample, can be found in \cite{bonPD97}.
It is enough to say that a certain semiinfinite problem
has to be solved numerically and matched to the
resulting situation with a narrow high-field domain
(consisting of a region of $E= E_3(J)$ bounded by
positively and negatively charged wavefronts) near
$x=0$ and a high-field region from a positively charged
wavefront to $x=L$. In the new situation, we have
$dJ/ds = A(J)\, [2\, c_+(J) -c_-(J)]$, and $J$ tries to
go toward the zero $J=J^{+}$ of $2\, c_+(J) -
c_-(J)$. Depending on the resistivity $\rho$, $J_c$ can
be larger than $J^{+}$ (and then $J$ decreases
toward $J^{+}$), or $J_c \in (J^{*},J^{+})$
(and then $J$ starts increasing, and a second
high-field domain may be expelled from $x=0$). The
simplest case, $J_c > J^{+}$, was described
asymptotically in \cite{bonPD97}. Provided $\phi$ is
large enough, $J$ evolves exponentially fast towards
$J^{+}$. When the old domain leaves the sample,
only two wavefronts (bounding the new high-field
domain) remain, and $J$ evolves exponentially fast
toward $J^*$, so that a period of the oscillation is
completed; see Fig.\ \ref{Ext.p1}. The second case, $J_c
\in (J^{*}, J^{+})$ is more complicated: numerical
simulations show that multiple high-field domains may
coexist in the sample at the same time as in Fig.\
\ref{Ext.p2}.

To achieve a simplified description of the current
oscillation, valid for any positive value of $J_c$, we
proceed to examine further the shedding process. The
simplest rule to determine when a new wavefront is
shed from the injecting contact would be as follows: a
positive (negative) front is emitted at the instant
that $J$ passes through $J_c$ with positive (negative)
time derivative. However, this rule neglects the time
needed for sufficient charge to be injected at the
contact to form a propagating front. We may estimate
the effective delay time by considering the time
evolution of Eq.\ (\ref{red}) evaluated at the
injecting contact,
\begin{eqnarray}
\dot{u} + J\, \frac{K + R}{V}\, u &=&
\left(\frac{1}{V} - \rho J\frac{V'}{V^{2}}
\right)\, \dot{J}\nonumber\\  &+& J\, \frac{K +
R}{V^{2}}\, [J - j(E)], \label{inj}
\end{eqnarray}
where $u(t)=\partial{E(0,t)}/\partial{x}$ and the
argument of $V$, $K$ and $R$ is $E(0,t) = \rho J(t)$, i.e., the
value of electric field at the injecting contact.
In this equation, we can think of $J(t)$ as driving
charge injection processes which determine front
launching. Based on extensive simulations of the
reduced model, we have found that $u(t)$ must attain a
sufficiently  large positive or negative value for the
front to detach and begin to propagate. This value is
generally found to lie between 50\%\ and 90\%\ of the
steady state value that $|u|$ would have in the case of
no propagating fronts, i.e., where $E$ near the
injecting contact rapidly rises to the $E_3$ value or
rapidly falls to the $E_1$ value. The asymptotic model
system is fully defined once the threshold is set and
consists in Eqs.\ (\ref{X_pm}) to (\ref{inj}). Thus, in
the limit of an infinitely long sample, terms of order
$\epsilon = 1/L$ drop and  we arrive at a
low-dimensional dynamical system, which consists
essentially of: (i)  propagating negative and positive
charge points that move according to the Eq.\
(\ref{X_pm}), (ii) subject to the conservation law Eq.\
(\ref{asy3}), (iii) which produce a measurable current
according to Eq.\ (\ref{asy2}), and are created
according to Eq.\ (\ref{inj}).  Note that in the
$\epsilon\to 0$ limit the current will exhibit slope
discontinuties at the formation times $t_{b,\pm}^{(i)}$ and
destruction times $t_{d}^{(i)}$, but will be otherwise continuous and
governed by Eq.\ (\ref{asy2}). We note that the
rigorous foundations of this and similar asymptotic
models have been explored recently using singular
perturbation methods \cite{bonPD97,bonPRE97}.

In this paper, we estimate the order-of-magnitude of
the time delay associated with wavefront formation by
evaluating Eq.\ (\ref{inj}) for $J\approx J_c$. This
implies a relaxation time of
\begin{equation}
\tau \approx\frac{V}{{dJ_{c}\over dt} - (\alpha-1)\, KV},
\end{equation}
i.e., the approximate time for $u(t)$ to
go from a value of $u(0)=-1$ to $u(\tau) = 0$. By
$dJ_c/dt$, we refer to the value of $dJ/dt$ when $J$
crosses $J_c$. Then, we adopt the criterion that a new
front is born at $x=0$ at the time $t+a\tau$ where $t$
is the time at which $J=J_c$, and $a\tau$ is a delay
time.  Here $a$ is a number of order one which is
determined from simulation of the reduced model for a
particular bias voltage and then assumed to apply over
the complete range. Wavefront destruction is assumed
occur instantaneously at times $t_{d}^{(i)}$ when
$X_{-}^{(i)} = X_{+}^{(i-1)}$ or when $X_{+}^{(i)} =
X_{-}^{(i)} = L$.  We ignore the finite duration of
(fast) monopole destruction stages which is equivalent
to the well-justified approximation of neglecting the
diffusive boundary layer at  the receiving contact
\cite{bonPD97}. It should also be kept in mind that the
index instantaneously decreases by one when wavefronts
downstream collide with one another.

Let us now use the above asymptotic model to interpret
the simulation results for Eqs.\ (\ref{red}) -
(\ref{bc}). The case of contact resistivity such that
$J_c > J^{+}$
 has been
explained already: we obtain the usual Gunn effect
with at most one solitary wave detached from the
contacts for any time \cite{bonPD97,bonPRE97}; see
Fig.\ \ref{Ext.p1}. Let us assume now that the contact
resistivity is such that $J_c \in (J^*,J^{+})$. Then
the current will increase after creation of a solitary
wave, because $2c_+ - c_- > 0$ and multiple wave
shedding is  possible \cite{bonPD97}. This situation is
shown in Fig.\ \ref{Ext.p22}, which shows simulation
results of our simplified asymptotic model for similar
parameters to those of Figure \ref{Ext.p2}. The latter
is depicted using data from direct numerical simulation
of the reduced {\em pde} model Eqs.\ (\ref{red}) - (\ref{bc}).
To obtain Fig.\ \ref{Ext.p22}, the values of $a$ were
set to 13.75 for positive front emission and 7.82 for
negative front emission. We use these same values in
the data of Fig.\ \ref{Ext.ch2}, which has same bias
values as Fig.\ \ref{Ext.ch1}. The chaos appears to be
closely tied to the asynchronous emission of fronts.
This explains why the chaos observed for this partial
differential equation system is low-dimensional. It is
interesting to speculate how the maximal number of
domains possible might scale with system size and
contact resistivity. 

\section{Conclusion}

We have utilized asymptotic analysis of a {\it pde} model (which describes the 
trap-dominated slow Gunn effect in a long sample) to explain the dynamics of 
space charge waves and current vs.\ time, including low-dimensional chaos which 
is nonetheless accompanied by spatially complex structure suggesting a loss of 
spatial coherence. The building blocks of this analysis are the heteroclinic 
orbits used to construct the typical solitary waves mediating Gunn-like 
oscillations. During most of the oscillation, the motion of the heteroclinic 
orbits and the change of the electric field inside and outside the solitary 
waves (enclosed by heteroclinic orbits) follow adiabatically the  evolution of 
the total current density. When a solitary wave reaches the receiving contact, 
the current increases abruptly and the asymptotic model adequately approximates 
this as instantaneous.  As an outcome, we have found a criterion that shows 
that single or multiple wave shedding is possible during each oscillation, 
depending on the resistivity of the injecting contact. While single shedding is 
the usual (stable) Gunn effect, multiple wave shedding may break the spatial 
coherence of the electric field within the sample. This new instability 
mechanism provides an explanation for the complicated behavior observed in 
experiments performed in long  semiconductor samples \cite{gun64,kahn45PRB92} 
and in the numerical simulation of the drift-diffusion model. We have confirmed 
these results by direct numerical simulation of the reduced model, in 
particular the new predictions of multiple shedding of solitary waves in the 
unstable case.  Although simulations and analyses have been presented here for 
the p-Ge model, the general approach is quite general and applies to a wide 
class of {\it pde} models which possess the following common properties: 1) an 
integral (over space) constraint; 2) standard boundary conditions which permit 
multiple stationary states - i.e., negative differential 
resistance; and 3) solitary waves (i.e., pulses) and fronts. It is interesting to 
speculate that different models may lead to the same class of long-sample 
asymptotic limiting model which completely determines and explains the long 
time dynamics (including chaotic temporal behavior and loss of spatial 
coherence) of the respective full models. 

It is a pleasure to acknowledge beneficial
conversations with P.\ J.\ Hernando and M.\ Kindelan.
We acknowledge financial support from the Spanish DGES
through grant PB98-0142-C04-01, and one of us (SWT) 
acknowledges support of the Fulbright Foundation.

\bibliographystyle{prsty}

\begin{figure}  
\caption{Stationary homogeneous
current density, $j(E)$, and linear contact
characteristic for a sample with the contact resistivity,
$\rho=$780 $\Omega$ cm
and compensation ratio $\alpha$=(acceptor
concentration)/donor concentration)=1.21.
}
\label{j_sh}
\end{figure}

\begin{figure}   
\caption{Space-time evolution of the electric
field $E(x,t)$ and the corresponding current density $J(t)$,
with parameter values
$\phi=6.25$ V/cm and $\rho = 780$ $\Omega$ cm. The gray scale
ranges from 5.3 V/cm (black) to 14.0 V/cm (white).}
\label{Ext.p1}
\end{figure}

\begin{figure}   
\caption{Space-time evolution of the electric field and
the corresponding current for $\phi=7.25$ V/cm.
}
\label{Ext.p2}
\end{figure}

\begin{figure}  
\caption{Space-time evolution of the electric field and the
corresponding current for a chaotic state with
$\phi=10.0$ V/cm and $\rho = 780$ $\Omega$ cm. }
\label{Ext.ch1}
\end{figure}

\begin{figure}   
\caption{ DC bias bifurcation diagram (a) local maxima
in the current and (b) largest dimensionless Lyapunov exponent.}
\label{phase}
\end{figure}

\begin{figure}  
\caption{This figure shows the velocities of the heteroclinic
orbits between $E_1(J)$ and $E_3(J)$, $c_+$ and
$E_3(J)$ and $E_1(J)$, $c_-$ vs. J, both of them
in dimensionless units. }
\label{c(J)}
\end{figure}

\begin{figure}  
\caption{Space-time evolution of the electric field and
the corresponding current density determined from the
asymptotic model for the period-2 state with $\phi=7.25$
V/cm and $J_c=9.983$ $\rm{mA/cm^2}$. }
\label{Ext.p22}
\end{figure}

\begin{figure}  
\caption{Space-time evolution of the electric field and
the corresponding current density determined from the
asymptotic model and showing a chaotic state for
parameters
 $J_c=9.983$ $\rm{mA/cm^2}$ and $\phi=10.0$ V/cm.}
\label{Ext.ch2}
\end{figure}

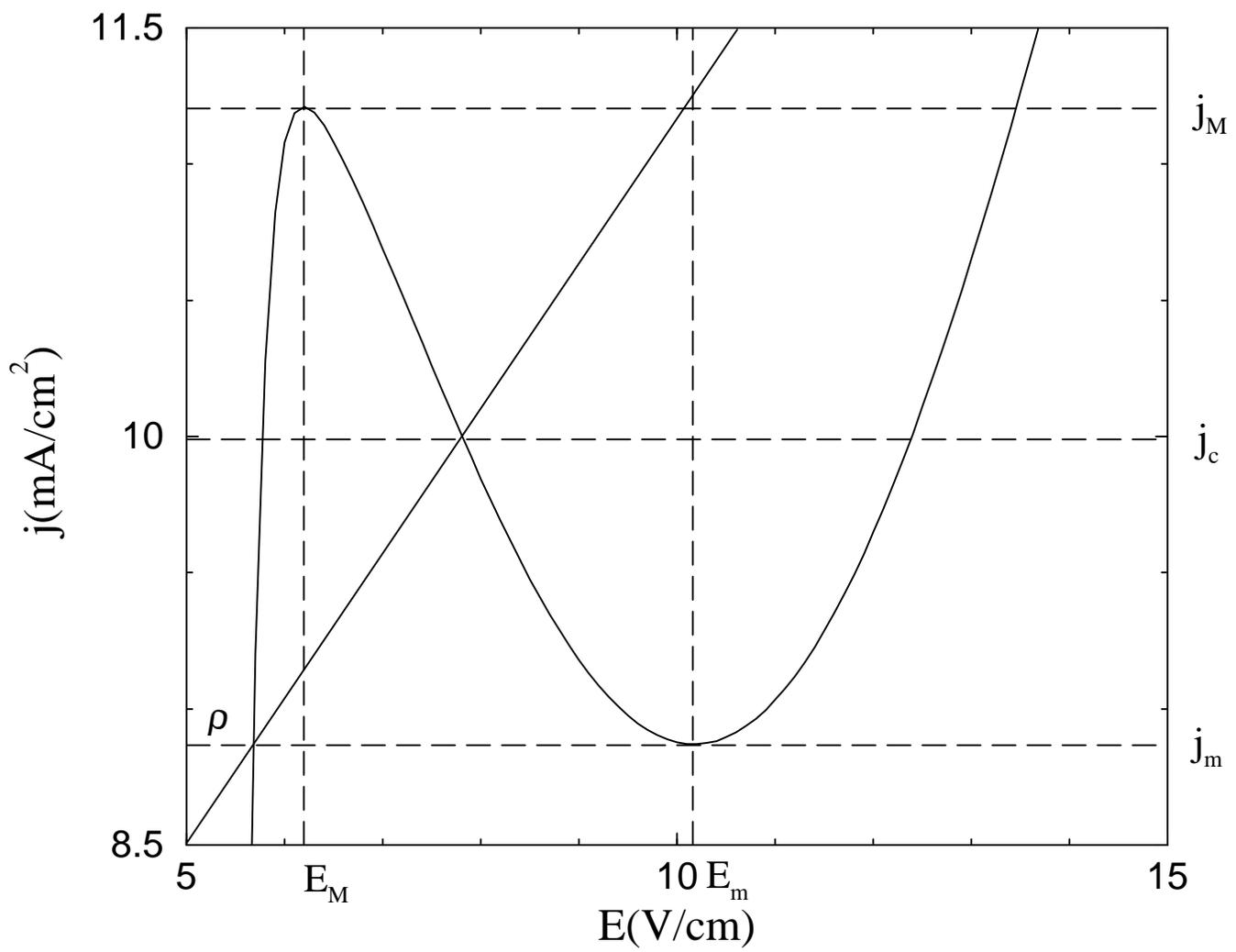
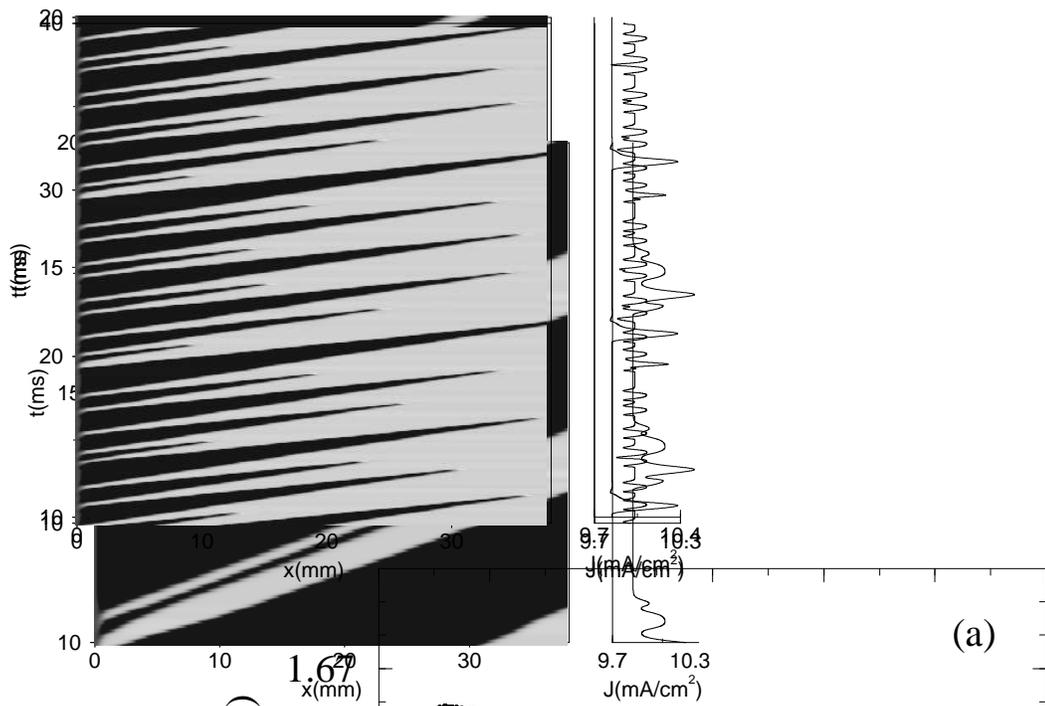
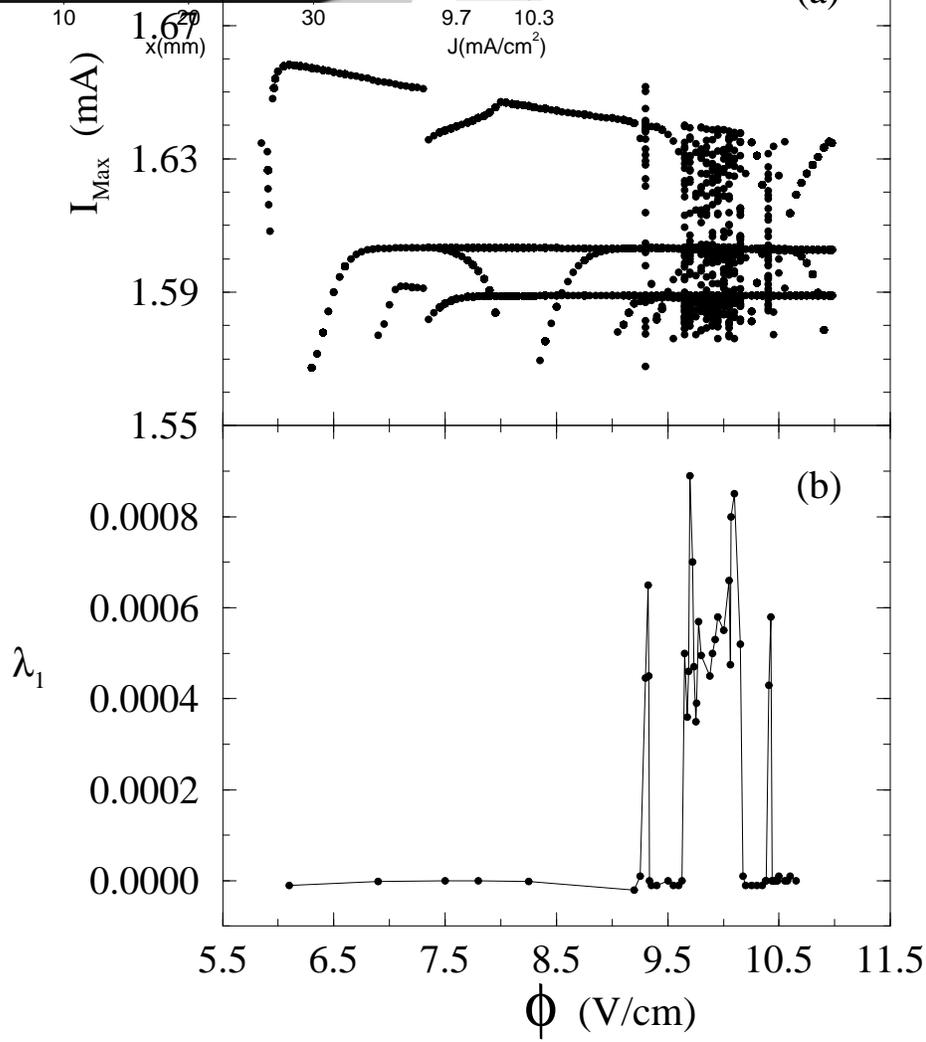
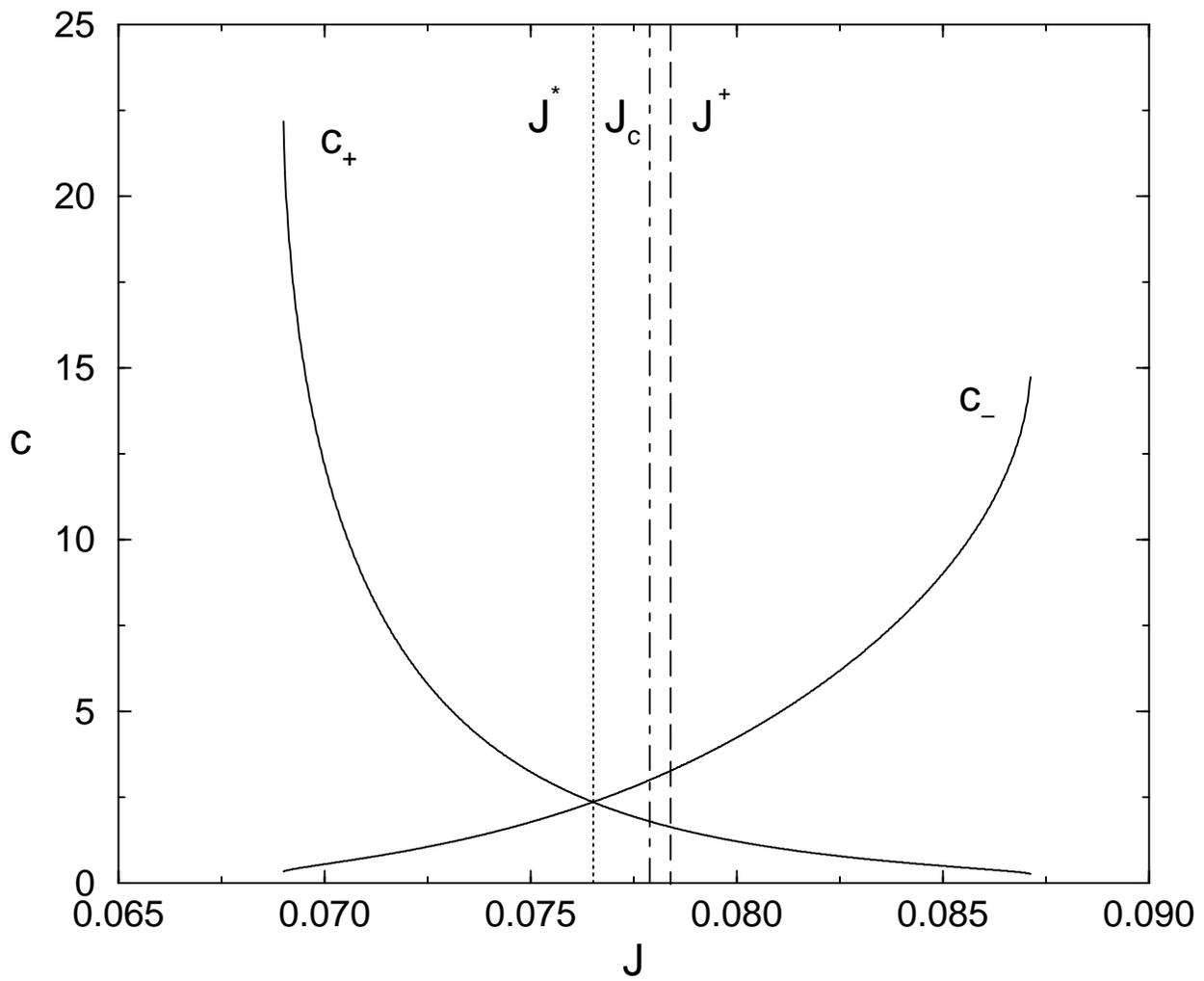
\end{document}